\documentclass[12pt]{iopart}

\usepackage{bm,amsfonts,amssymb, graphicx}
\usepackage{cite}
\usepackage{epstopdf} 
\begin{document}

\title[Multiphoton excitation and high harmonic generation in RGQD]{Multiphoton excitation and high harmonic generation in rectangular graphene quantum dot}

\author{A G Ghazaryan$^{*} $ and Kh V Sedrakian}

\address{Centre of Strong Fields Physics, Yerevan State University, 1 A. Manukian,
Yerevan 0025, Armenia\\}
\ead{amarkos@ysu.am}
\vspace{10pt}
\begin{indented}
\item[{\today } ]
\end{indented}

\begin{abstract}
The multiphoton excitation and high harmonic generation (HHG) processes are considered using the microscopic quantum theory of nonlinear interaction of strong
coherent electromagnetic (EM) radiation with rectangular graphene quantum dot (RGQD). 
The dynamic Hartree-Fock approximation is
developed for the consideration of the quantum dot-laser field
nonlinear interaction at the nonadiabatic multiphoton
excitation regime. The many-body Coulomb interaction is described in the extended
Hubbard approximation. By numerical results, we show the significance of the
RGQD lateral size, shape, and EM wavefield orientation in RGQD of the zigzag edge compear to the armchair edge in the HHG process 
allowing for increasing the cutoff photon energy and the quantum yield of higher harmonics. 
\end{abstract}

%\maketitle

\section{Introduction}

The quantum electrodynamic phenomena caused by the strong coherent EM radiation, such as multiphoton HHG and related processes, 
through nonlinear channels in two-dimensional (2D) atomic
systems-nanostructures are of great interest in lowenergy physics and nanooptoelectronics due to the unique physical properties of
such 2D nanosystems of atomic thickness \cite{1,1bb,2,hhg1,ww1,ww2,ww3,Abook,hhg2}. 
It was been studied systematically the HHG in bulk crystals \cite{s1,s2,s3,s5,s6,s7,s8} and low-dimensional 
nanostructures such as graphene and its derivatives \cite{H2,Mer,Mer1,H3,H4,H6,H7,H8,H9,H99, H10, H11, H12, H12a, H13, H14, H15, H16, H17, H18, H19, H20, H21}%
, monolayer transition metal dichalcogenides \cite{TMD,TMD1,TMD11}, 
hexagonal boron nitride \cite{BN}, topological insulator \cite{TI}, \cite{TII}, monolayers of black
phosphorus \cite{phosph}, curved 2D hexagonal nanostructures \cite{Mer2019}, 
solids \cite{corcumsolid}, \cite{semimetal}, as well as in other 2D systems \cite{twodem1,twodem2,twodem3}. 
2D nanosystems enable the development of important 
technological applications \cite{cc1}. The quantum cascade laser is one such example \cite{QCL}, 
using physical phenomena in 2D systems such as the quantum Hall effect \cite{Abook}.

As a nonlinear medium, the graphene quantum dot (GQD) \cite{cc1}, 
\cite{1111, arxiv,arxiv1,arxiv2}, and such as graphene nanoribbon \cite{1}, \cite%
{bb,new1,new2,62,63,64,65,66,bb1,dd,GNR1,GNR2}, are of particular interest. These can be closed and convex structures such 
as fullerenes of various basic symmetry, as well as plane structures of various lateral sizes, 
shapes, edges, and doping levels \cite{bb,new1,new2}. As known, the symmetry 
of the graphene sublattice in GQD can be controlled by the lateral size, shape, and type of the GQD edge 
\cite{arxiv1,arxiv2,bb,new1,new2,62,63,64,65,66,bb1,dd,GNR1,GNR2,bb11,bbb}. The behavior of 
GQD is quantitatively different for nanostructures with zigzag and armchair edge \cite{ff}, \cite{ff1}. 
Thus, GQDs have richer electronic properties than graphene unbounded in space \cite%
{26,27,28}. Moreover, they are of particular interest because their nonlinear optical properties can be controlled \cite{1}. 

The 2D graphene unbounded in space formed in the type of a narrow ribbon leads to 
the confinement of carriers in quasi-one-dimensional graphene nanoribbons (with different topologies depending 
on the shape of the ribbon) \cite{H99}. An important advantage 
of GQD over graphene nanoribbons \cite{GNR1}, \cite{ee1,ee2,ee3} is the complete confinement 
of quasiparticles in space. For a bounded quantum system, due to the lack of translational symmetry, 
the spectrum only consists of a discrete set of energy levels, not bands. One of the advantages of 
limited quantum systems is the ability to change their energy spectra and wavefunctions by 
adjusting their size \cite{ee4}, \cite{ee5} or external parameters 
such as gate voltage or magnetic field \cite%
{ee6,ee7,ee8}. Note that the HHG spectra in GQDs can be affected by the 
constraint conditions — the lateral size of the dot and/or the coupling parameters in 
the bounded GQDs \cite{bb1}, \cite{arxiv1},  \cite{arxiv2}. The HHG efficiency is expected to 
increase with increasing confinement since the latter will limit the propagation of the electron wave packet \cite{Lew}.%
\begin{figure}[tbp]
\includegraphics[width=.7\textwidth]{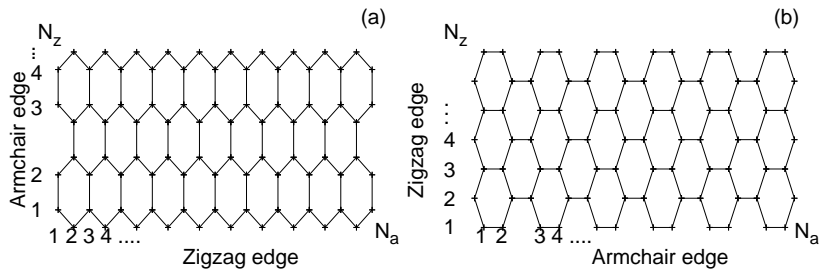}
\caption{ The geometrical structure of ($N_{a}$; $N_{z}$)-sized RGQD with $N=N_{a}N_{z}$ atoms. 
(a) shows RGQD with zigzag edge along the $x$ axis with $N=84$ atoms for $N_{a}=21$; $N_{z}=4$, 
(b) -- RGQD with armchair edge along the $x$ axis for $N_{a}=12$; $N_{z}=7$. 
The distance between nearest neighboring atoms is $b\simeq1.42\ \mathring{A}$.}
\end{figure} %

In the present paper, we consider the multiphoton HHG in the RGQD--plane 
quasi-zero-dimensional GQD of the rectangular shape 
\cite{GQD1,GQD2,GQD33}, with zigzag and armchair edges on the elongated side of the different lateral sizes induced
by intense coherent radiation in the nonlinear regime. The many-body
Coulomb interaction is taken into account in the extended Hubbard approach \cite{new1,new2},  \cite{Kast}. 
Note that the considered GQDs, where the engineering of optoelectronic properties is allowed, currently 
are accessible in practice \cite{bb}%
, \cite{dd}. A closed system of differential equations for a single-particle
density matrix at the multiphoton interaction of RGQD with a strong laser
field is solved numerically.

The paper is organized as follows. In Sec. II, we consider multiphoton
excitation and HHG in RGQD of the different lateral
sizes, laser intensity, and polarization of the wave
field. The comparison with the case of 
the armchair edge on the elonged side of RGQD was done. Conclusions are given in Sec. III.  Finally, in Sec. IV the set of equations for the
single-particle density matrix has been formulated, taking into account the
many-body Coulomb interaction. %

\section{Basic model, numerical results and discussions}

We will study in RGQD the HHG of the plane quasi-monochromatic EM wave of linear polarization, propagating perpendicular to the $xy$ plane, 
with homogeneous quasi-periodic electric field strenth:%
\begin{equation}
\mathbf{E}\left( t\right)=\widehat{\mathbf{e}} E_{0}f\left( t\right)\cos\omega t,  \label{el}
\end{equation}% 
where $ \widehat{\mathbf{e}}$ is the unit polarization vector in the $xy$ plane,
 $E_{0}$ and $\omega $ respectivelly are the amplitude and carrier frequency,
$f\left( t\right) =\sin ^{2}\left( \pi t/\mathcal{T}\right) $ is the slowly changing envelope, $%
\mathcal{T}=40\pi /\omega $ is the impulse duration. Fig. 1 shows the geometric structure of the zero-dimensional  RGQD in $xy$ plane. There are $N_{a}$ and $N_{z}$
 carbon atoms along the $x$ axis and the $y$ axis, respectively. All carbon atoms are 
packed into a hexagonal lattice. The total number of carbon atoms RGQD is $N=N_{a}N_{z}$. 
Note that, for nanoribbons with a limited armchair edge width, the electronic structure and 
energy spectrum of quasiparticles for
 $N_{a}>>N_{z}$ and $N_{a}<<N_{z }$, when the RGQD will be similar to a nanoribbon, critically depend on the width of 
the RGQD along the $y$ axis and the width along the $x$ axis, respectively \cite{GNR2}. 
In relation to \cite{GNR2}, we have the insulator case of RGQD (Fig. 1).%
\begin{figure}[tbp]
\includegraphics[width=.62\textwidth]{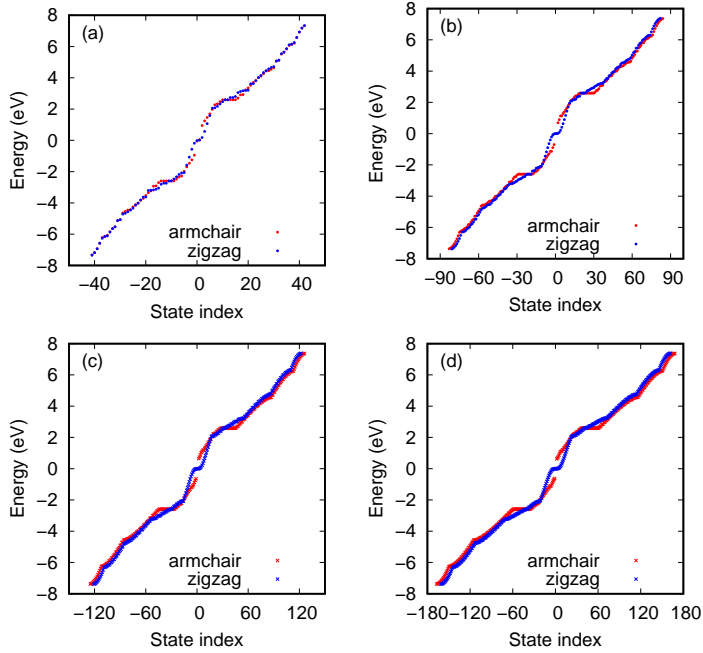}
\caption{ (Color online) The eigenenergies in RGQD. 
(a--d) for zigzag edge along the $x$ axis respectively correspond to 
$N=84 (N_{a}=21), 164 (N_{a}=41), 244 (N_{a}=61), 324 (N_{a}=81)$, atoms; and $N_{z}=4$ for all cases.
The same for RGQD of armchair edge  along the $x$ axis for (a--d) 
$N=84 (N_{a}=12), 168 (N_{a}=24), 252 (N_{a}=36), 336 (N_{a}=48)$ atoms, respectively; and $N_{z}=7$ for all cases.}
\end{figure} %

Thus, we will use the microscopic nonlinear quantum theory of the HHG process in the strong EM wave in RGQD by the 
tight-binding model (TB) \cite{bb}, \cite{Wal} for the Hamiltonian $ \widehat{H}$, 
taking into account the Coulomb interaction in the generalized Hubbard approximation. The total Hamiltonian by the empirical
 TB model \cite{Wal} is given in Sec. IV. Using numerical diagonalization, we will find the eigenstates $\psi _{\mu }\left( i\right) $ and
eigenenergies $\varepsilon _{\mu }$ ($\mu =0,1..N-1$). 
The results of numerical diagonalization are shown in Fig. 2 for the different lateral
sizes and edges of RGQD. As shown in Fig. 2, without the tunneling all energy levels have degenerated. 
It can also be seen from Fig. 2 that the density of states increases with an increase in the number of lattice atoms.
 As will be seen later, this directly affects the yield of HHG.

The basic theory is given in Seq. Appendixes. The quantum dynamics of RGQD in a strong EM wave 
with periodic excitation is determined by a closed system 
of differential equations (\ref{evEqs}) for the density matrix, which must be solved with the certain initial conditions. We construct mainly the 
density matrix $\rho_{0ij}^{\left( \sigma \right)}$ through the filling of electronic states in
valence band in accordance with the Fermi--Dirac distribution at zero temperature %
\begin{equation}
\rho _{0ij}^{\left( \sigma \right) }=\sum_{\mu =N/2}^{N-1}\psi _{\mu }^{\ast
}\left( j\right) \psi _{\mu }\left( i\right),  \label{dens}
\end{equation}% 
 with an eigenstate $\psi _{\mu }\left( i\right)$ of the Hamiltonian $\widehat{H}_{0}$ (\ref{Hfree}). For the density 
matrix we solve numerically by integration the equations of motion Eq. \ref{evEqs} 
over time with the standard fourth-order Runge-Kutta algorithm.

The emission spectrum of harmonics is defined by the Fourier transform 
$\mathbf{a}\left( \Omega \right) $ of the dipole acceleration:%
\begin{equation}
\mathbf{a}\left(t\right) =d^{2}\mathbf {d}/dt^ {2}.  \label{dens}
\end{equation}% 
The dipole momentum is:%
\begin{equation}
\mathbf{d}\left(t\right) =\left\langle \sum_{i\sigma}\mathbf{r}_{i}c_{i\sigma}^{\dagger}c_{ i\sigma}\right\rangle,  \label{dipol}
\end{equation}% 
normalized by the number $N$ of lattice atoms, and the factor $a_{0}=\overline{\omega }^{2}{d}_{0},$ where $\overline{%
\omega }=1\ \mathrm{eV}/\hbar $ and ${d}_{0}=1\ \mathrm{\mathring{A}}$. 
The power emitted at a given frequency is proportional to $\left\vert 
\mathbf{a}\left( \Omega \right) \right\vert ^{2}$. We introduce the angle $\theta $ between the polarization vector 
$\widehat{\mathbf{e}}=\left\{ \cos \theta ,\sin \theta \right\} $ and the $x$ 
axis directed by the alonged edge (see Fig. 1). The relaxation rate is taken $\hbar \gamma =50\ \mathrm{%
meV}$. The electron-electron interaction (EEI) energy of on-site Coulomb repulsion is $U\simeq 3\ \mathrm{eV}$ (see in Sec. IV). 
The inter-site Coulomb repulsion energy is taken $V\simeq 0.9\ \mathrm{eV}$. The energy between
the nearest-neighbor atoms is $t_{ij}=2.7\ \mathrm{eV}$. 
To clarify the main aspects of multiphoton HHG in RGQD, we assume that the excitation 
frequency is $\omega =0.1\ \mathrm{eV}/\hbar $, which is much smaller than the typical scales 
$t_{ij}$, $U$, $V$. %
\begin{figure}[tbp]
\includegraphics[width=.62\textwidth]{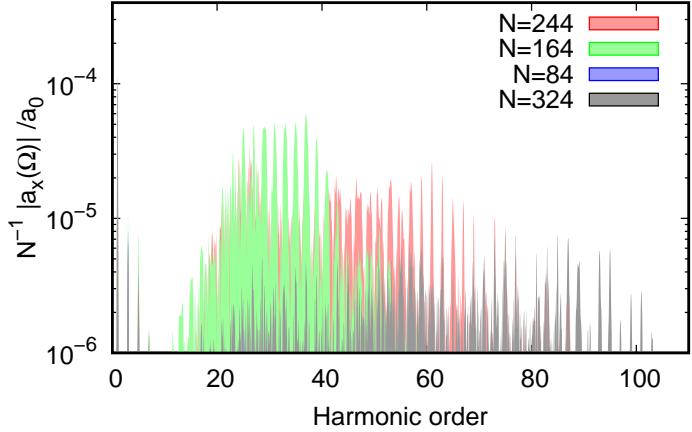}
\caption{(Color online) The HHG emission spectra in the strong-field regime
via dipole acceleration Fourier transformation $%
N^{-1}|a_{x}\left( \Omega \right) |/a_{0}$ (in arbitrary units) versus the
harmonic number for RGQD with zigzag edge on the elongated side of different number $N$ of atoms. The relaxation
rate is $\hbar \protect\gamma =50\ \mathrm{meV}$. The linearly polarized
wave frequency is $\protect\omega =0.1\ \mathrm{eV}/\hbar $ and wavefield
strength is $E_{0}=0.1\ \mathrm{V/\mathring{A}}$  at fixed
angle $\protect\theta =0$. The spectra are shown for
moderate typical EEI energies: $U=3\ \mathrm{eV}$, $V=0.9\ \mathrm{eV}$. }
\end{figure}%
\begin{figure}[tbp]
\includegraphics[width=.62\textwidth]{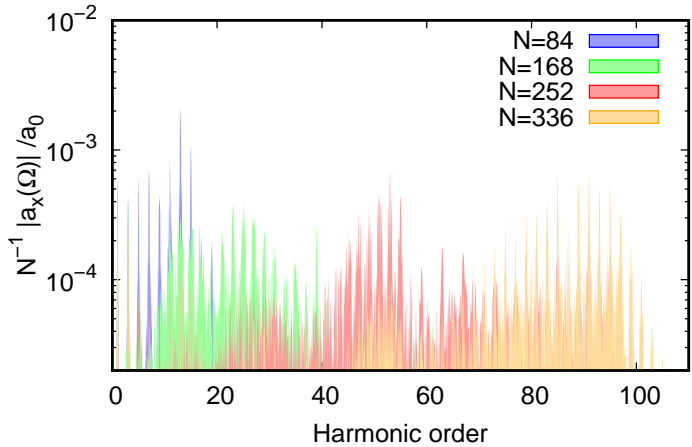}
\caption{(Color online) The same as for Fig. 3 but for RGQD of armchair edge  along the $x$ axis.}
\end{figure}%

Next, we will consider the origin of the HHG in the quantum dot. There are two contributions to the current: 
the electron/hole transitions within unoccupied/occupied states and the electron-hole 
creation (transitions from occupied states to unoccupied ones) and
subsequent recombination. The first transitions make a contribution only to 
low harmonics and is analogous to the intraband current in a semiconductor, 
while the latter makes the main contribution in the high-frequency part corresponding 
to the interband current, which represents recombination/creation of electron-hole pairs. 
This picture is analogous to HHG in solid-state systems. To separate these contributions in 
the dipole acceleration spectrum we made a change of the basis via formula $\rho
_{ij}=\sum_{\mu ^{\prime }}\sum_{\mu }\psi _{\mu ^{\prime }}^{\ast }\left(
j\right) \rho _{\mu \mu ^{\prime }}\psi _{\mu }\left( i\right) $, where $%
\rho _{\mu \mu ^{\prime }}$ is the density matrix in the energetic
representation. Hence intraband part of dipole acceleration defines  as %
\begin{equation}
\mathbf{d}_{\mathrm{intra}}\left( t\right) =\sum_{\mu ,\mu ^{\prime
}=N/2}^{N-1}\rho _{\mu \mu ^{\prime }}\left( t\right) \mathbf{d}_{\mu
^{\prime }\mu }+\sum_{\mu ,\mu =0}^{N/2-1}\rho _{\mu \mu ^{\prime }}\left(
t\right) \mathbf{d}_{\mu ^{\prime }\mu },  \label{dintra}
\end{equation}%
and interband part will be %
\begin{equation}
\mathbf{d}_{\mathrm{inter}}\left( t\right) =2\sum_{\mu ^{\prime
}=N/2}^{N-1}\sum_{\mu =0}^{N/2-1} Re \rho _{\mu \mu ^{\prime }}\left(
t\right) \mathbf{d}_{\mu ^{\prime }\mu },  \label{dinter}
\end{equation}%
where dipole transition matrix elements are: %
\begin{equation}
\mathbf{d}_{\mu ^{\prime }\mu }=\sum_{i}\psi _{\mu ^{\prime }}^{\ast }\left(
i\right) \mathbf{r}_{i}\psi _{\mu }\left( i\right).  \label{moment}
 \end{equation}%
\begin{figure}[tbp]
\includegraphics[width=.6\textwidth]{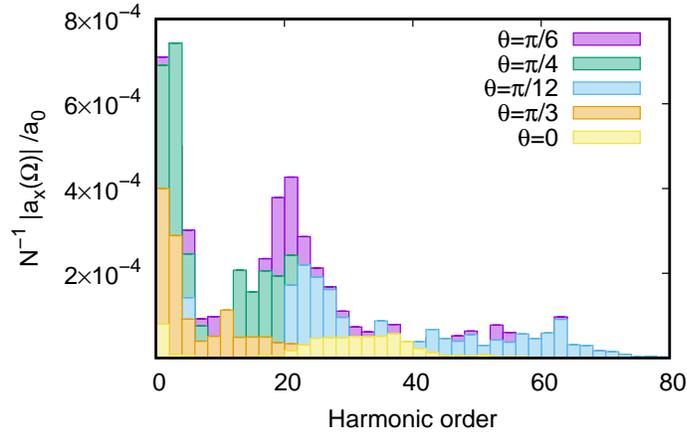}
\caption{(Color online) The same as for Fig. 3 but for different angles $%
\protect\theta $ composed between in-plane electric field and $x$ axis for $%
N=164$ atoms of RGQD.}%
\end{figure}%
\begin{figure}[tbp]
\includegraphics[width=.6\textwidth]{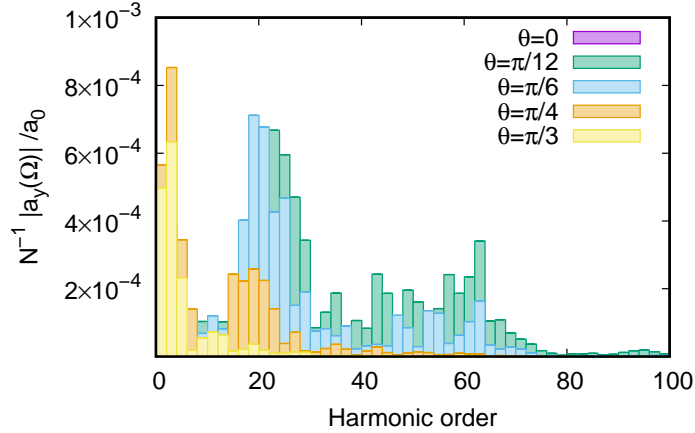}
\caption{(Color online) The same as for Fig. 5 but for $a_{y}$--component.}
\end{figure}% 
In the following we will consider two contributions 
intraband  (\ref{dintra}) and interband (\ref{dinter}),  to the total current numerically.

In Figs. 3, 4 the HHG spectrum determined by $|a_{x}\left(\Omega \right) |$ is presented in RGQD of the different lateral
sizes of correpondently zigzag and armchair edge on the elonged side. The
 linearly polarized EM wavefield is taken at polarization angle $\theta =0,$ % 
$E_{0}=0.1$ V$\mathrm{/\mathring{A},}$ $\omega = 0.1\ \mathrm{eV}/\hbar $. 
The EEI energies are $U\simeq 3\ \mathrm{eV}$, $V\simeq 0.9\ 
\mathrm{eV}$. As expected, the appearance of odd harmonics in the HHG spectrum is associated with the 
conservation of inversion symmetry in RGQD, as in ordinary graphene \cite%
{1}, \cite{arxiv}. This is a result of the interference of the two different contributions, 
intraband  (\ref{dintra}) and interband (\ref{dinter}). In Figs. 3, 4 we see the typical nonperturbative 
behavior of HHG spectra with a multiple plateau structure. 
In these cases, the dominant plateau shifted towards higher frequencies with 
an increase in the number of atoms. In particular, as shown in Fig. 3, at $N=164$, $N=244$, and $N=324$ we have an 
effective generation of harmonics from the 10th to the 60th, the 20th to the 90th, and from the 20th to the 100th, 
respectively. When, as shown in Fig. 4, in RGQD of armchair edge in RGQD,
 in particular for $N=168$ and $N=336$, we have an
efficient generation of harmonics from 10th to 40th and 70th to 100th,
respectively. This is related to the energy spectrums in Fig. 2, in which, with an increase in the 
number of lattice atoms, new higher energy states with an increasing density of states appear. 
As shown in Figs. 3 and 4, there are main differences in the higher harmonics spectra 
for the close lateral size RGQDs of different edges on the elonged sides. Thus, 
the HHG spectra maxima and cutoff energies are different in RGQDs of zigzag and armchair edges on the elonged sides. As will be shown by the following (see Fig. 14, red bar), for RGQD 
at some angles $\theta >0$ the magnitudes of high-order harmonics in the elonged zigzag case can 
become larger than at the armchair edge case, especially for $a_{y}$. %
\begin{figure}[tbp]
\includegraphics[width=.6\textwidth]{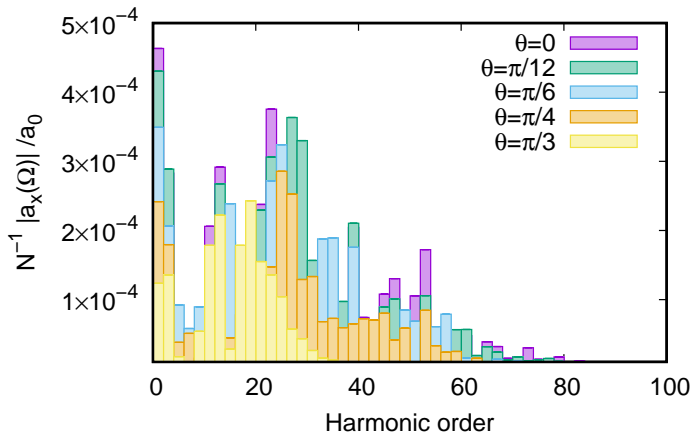}
\caption{(Color online) The same as for Fig. 5 but for $%
N=168$ atoms of RGQD of armchair edge along the $x$ axis.}%
\end{figure}%
\begin{figure}[tbp]
\includegraphics[width=.6\textwidth]{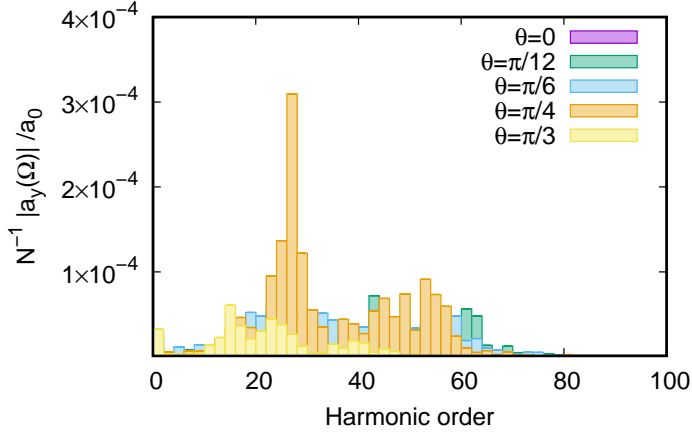}
\caption{(Color online) The same as for Fig. 7 but for $a_{y}$--component.}
\end{figure}% 

Furthermore, RGQD also has a strong anisotropic nonlinear response, depending on the orientation of 
the EM wavefield relative to the elonged side \cite%
{arxiv1},  \cite{arxiv2}. For RGQD at $\theta =0$ and $\theta =\pi /2$, the harmonic polarization 
direction coincides with the incident wave polarization direction. At other angles $0<\theta <\pi /2$, 
harmonics appear with polarization vectors perpendicular to the pump wavefield. To reveal the dependence of the HHG spectra on the orientation of the EM wavefield in Figs. 5 and 6 
show the HHG spectra for different $\theta$ of the pump wavefield relative to 
the $x$ axis in the RGQD of zigzag edge for $N=164$, for $|a_{x}|$ and $|a_{y}|$ components, respectively. 
In contradistinction to the isotropic HHG spectrum for unbounded in space graphene at the low-frequency pump wave, the spectrum 
is anisotropic at a high-frequency wave, when, due to the symmetry of the carbon hexagonal cell, 
the optical response with respect to the driven field polarization is periodic with a period of $\pi/3$ \cite{polar}. 
For the RGQD we have strong anisotropy. In particular, for HHG process in the case of more than the 
first eight harmonics for the $y$ component the angles $ \pi  /12\lesssim \theta \lesssim \pi /6$ are preferable, while for 
the $x$-component we have a maximum for the angle $\pi /6$. Moreover, different polarization angles lead to different maxima in the harmonic spectra 
and cutoff energies. In Figs. 7, 8 the same is shown for  RGQD of armchair edge.
For this case we see a completely different picture.  In
particular, for the x-component the angles $0\lesssim \theta \lesssim \pi
/12 $ are preferable, meanwhile for the $y$-component we have a maximum near
the angle $\theta \sim \pi /4$. The latter is also a 
consequent of the rich spectra of eigenstates in Fig. 2 with different symmetry. Note that during 
the generation process, only odd harmonics appear in the RGQD, regardless of its orientation. 
This is due to the inversion symmetry of the sublattice of RGQD. %
\begin{figure}[tbp]
\includegraphics[width=.9\textwidth]{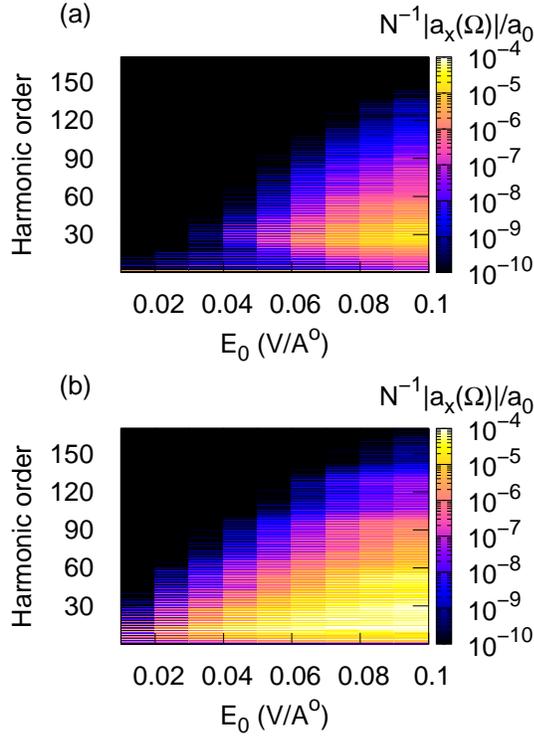}
\caption{(Color online) The color bars represent the HHG emission rate in
strong field regime in the logarithmic scale via dipole acceleration Fourier
transformation $N^{-1}|a_{x}\left( \Omega \right) |/a_{0}$ (in arbitrary
units) versus harmonic order and the wavefield strength $E_{0}$, at fixed
angle $\protect\theta =0$ for (a) $N=164$ at zigzag edge and (b) $N=168$ at armchair edge of RGQD on the elonged side. The wave
frequency is $\protect\omega =0.1\ \mathrm{eV}/\hbar $ and the EEI energies
are $U=3\ \mathrm{eV}$ and $V\simeq 0.9\ \mathrm{eV}$. The relaxation rate
is $\hbar \protect\gamma =50\ \mathrm{meV}$.}%
\end{figure}%
\begin{figure}[tbp]
\includegraphics[width=.92\textwidth]{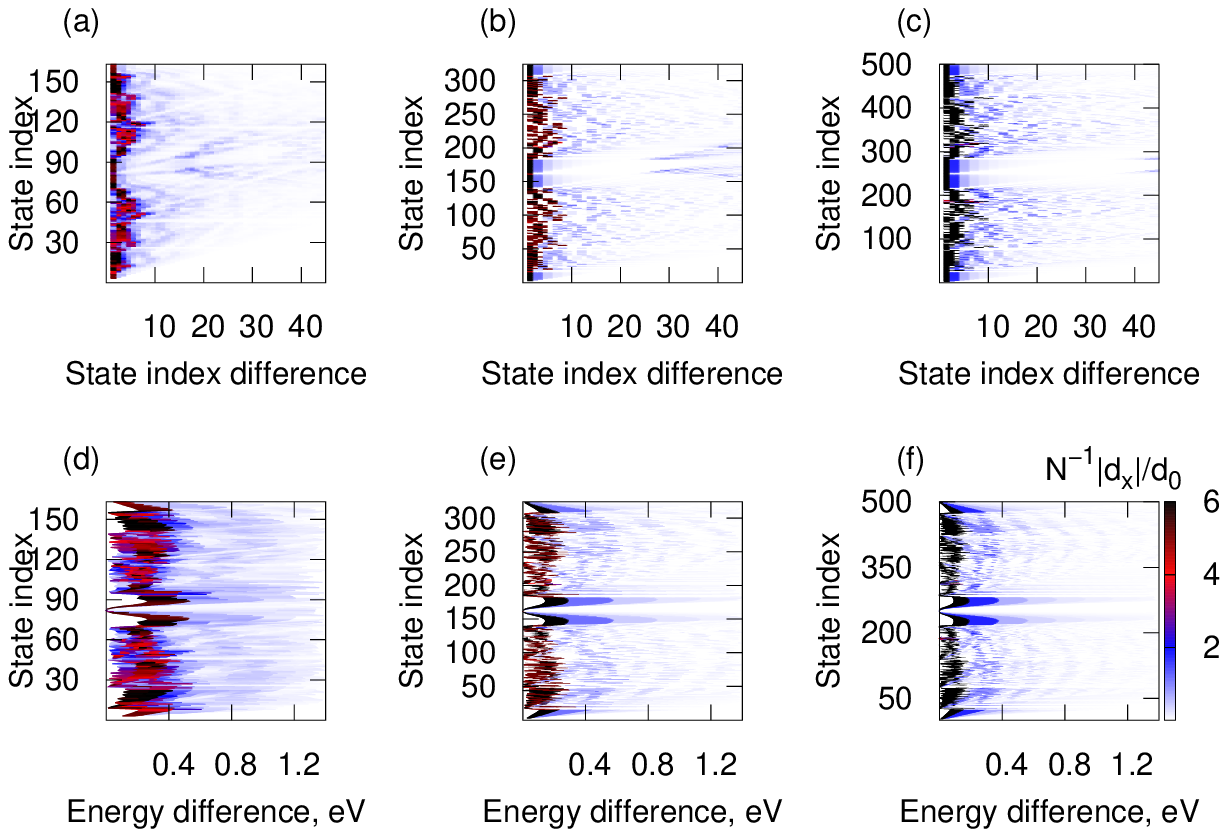}
\caption{(Color online)  The color bars represent the dipole momenta in the logarithmic scale $N^{-1}|d_{x} |/d_{0}$ (in arbitrary
units) versus eigenstate index difference and state index.
(a--c) respectively correspond to $N=164, 324, 500$ atoms of RGQD with a zigzag edge along the $x$ axis.
(d--f) show the same but versus eigenstate energy difference and state index.}
\end{figure}%

Next, we consider the HHG spectra depending on the intensity of the pump wave. 
Fig. 9 (a) shows the HHG spectra via the field strength and 
the harmonic order for a fixed frequency and EEI energies $U=3$ $\mathrm{eV}$, $V\simeq 0.9\ \mathrm{eV}$ 
in RGQD of zigzag elonged edge with $N=164$ atoms. We will mention, that for HHG it is important to increase the HHG emission efficiency and
 increase the harmonic cutoff threshold.  As shown in Fig. 9 (a), within each plateau, the 
cutoff harmonic increases linearly with increasing wavefield strength. 
Then, reaching harmonics $n_{\mathrm{cut}}\simeq 110$. Note that the linear dependence of the cutoff harmonics on
the wave intensity is similar to HHG through discrete levels \cite{prb,pra,46}, or in crystals with linear energy 
dispersion \cite{TII}, \cite{Vamp1}. To compear in Fig. 9 (b) we show the same for RGQD of the armchair edge on the elongated side. 
Fig. 9 has demonstrated the same multi-plateau behavior. Many differences appear in Fig. 9 (a) and (b), in particular, 
the absence of the rates of harmonic orders less $n=30$ in Fig. 9 (a), depending 
on the dipole momentum transition matrix elements (\ref{dipol}) in the RGQD of 
the zigzag or armchair edges are defined by the internal static property of the system of the quantum dot.

To reveal such differences via edge on the elongated side of the RGQD for different lateral
sizes, in Figs. 10 and 11 was shown the $x$--component of the dipole momentum $|\mathbf{d}_{\mu ^{\prime }\mu }|$ (\ref{moment}) 
via the eigenstate index difference and energy difference (transition energy), correspondently demonstrated for a zigzag and armchair edges. %
\begin{figure}[tbp]
\includegraphics[width=.92\textwidth]{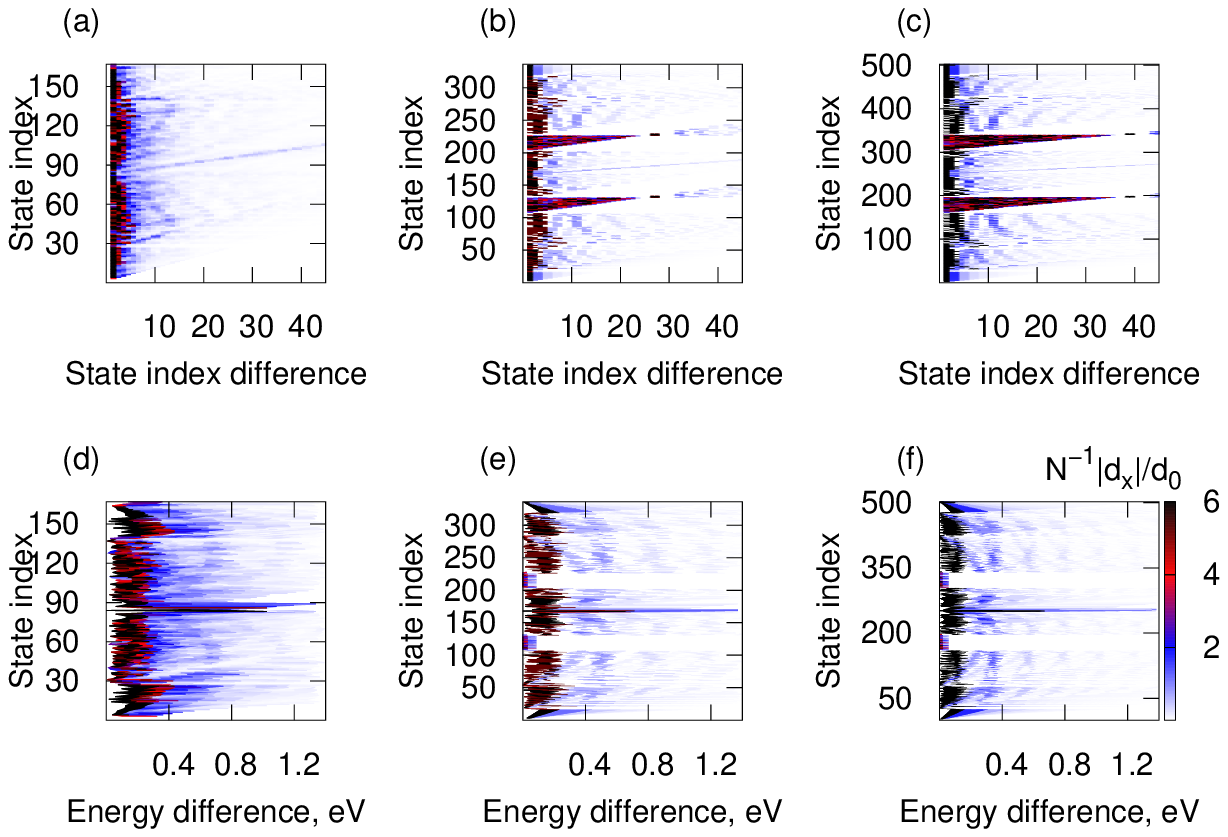}
\caption{(Color online) The same as for Fig. 10  but for RGQD with an armchair edge for (a--c), as for as (d--f), corresponding to $N=168, 336, 504$ atoms, respectively.}
\end{figure}%
 In Figs. 10, 11 appear, that with a change in the 
state index, new higher energy states with an increasing density of states appear, 
and the states are strongly different for zigzag and armchair edges of RGQD 
(connecting with different topologies depending on the edge on the elongated side of the quantum dot). As it is shown in Figs. 10, 11, 
with the growth in the number of carbon atoms 
the corresponding HHG rates related by the transitions with the more significant change of the state index 
occur relating to interband transitions, moreover, it prevails 
in the case of armchair edge. Despite this, as shown in Fig. 10, 11 for (c-f), in both cases of  RGQD edges, the energy difference of transitions tends 
to saturation with the increasing number of carbon atoms.

Is in the particular interest the dipole momentum $y$--component. The transitions, relating with $d_{y}$, 
appear mainly due to the strong coherent EM radiation. In Figs. 12, 13 $d_{y}$--components 
are shown for different edges and atom numbers of RGQD. 
As Figs. 12 and 13 show, that although the $d_{y}$--component by 
an order of magnitude is smaller than the $d_{x}$--component, the HHG rates are defined 
by correspondent transitions between eigenstates with higher index difference, than 
in the case of $d_{x}$--component; and for the RGQD of the armchair edge, the dipole momentum magnitude is larger 
than in the case of zigzag edge on the elonged side. In contradiction with Fig. 13 of the quantum dot of the elonged 
armchair edge, in Fig. 12 for the zigzag edge, the middle isolated lines corresponding to the substantial increase 
in the magnitude of $d_{y}$--component is clearly visible, which due to interband transitions (defined by 
respective state index difference).  The latter has the main contribution to the high-frequency 
part corresponding to the interband current, which is connected with recombination/creation 
of electron-hole pairs, and it will correspondently increase the HHG rates. Moreover, Figs. 12, 
13 for (d--f) show the satiation of the dipole momentum magnitude depending on the state energy difference with the atom number growth, as in Figs. 10, 11 for (d--f).%
\begin{figure}[tbp]
\includegraphics[width=.92\textwidth]{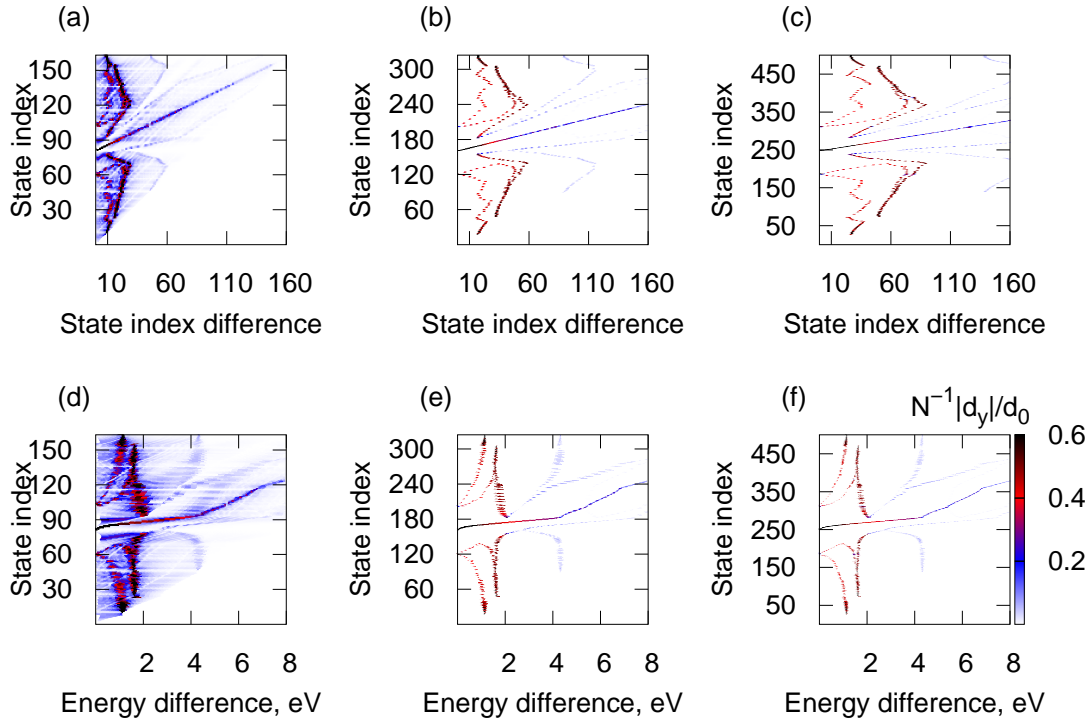}
\caption{(Color online) The same as for Fig. 10 but for $d_{y}$--component.  }
\end{figure}%
\begin{figure}[tbp]
\includegraphics[width=.92\textwidth]{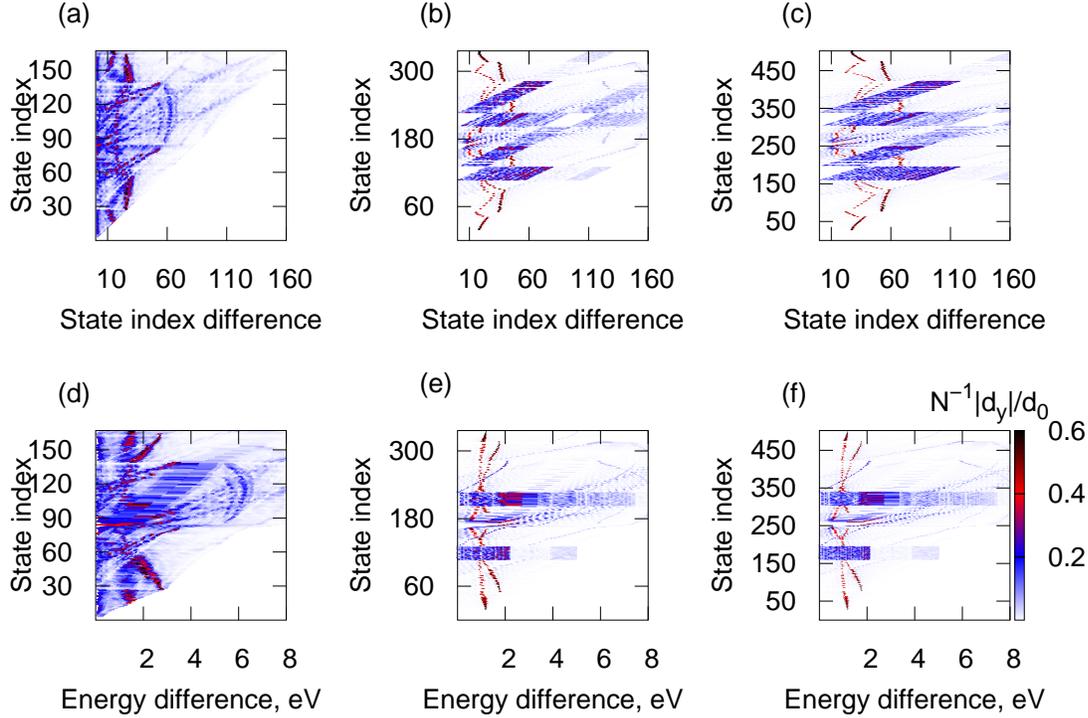}
\caption{(Color online) The same as for Fig. 11 but for $d_{y}$--component. }
\end{figure}%

In relation to the mentioned, Fig. 14 shows the high harmonic spectrum versus the angle $\theta$ of the EM wavefield 
relative to the elonged side, and harmonic order. (a) and (c) are demonstrated $|a_{x} |$ and 
$|a_{y} |$, respectively,  in RGQD of zigzag edge on the elonged side. (b) and (d) correspond $|a_{x}|$ and 
$|a_{y}|$, respectively,  in RGQD of armchair edge. Fig. 14 is demonstrated 
for $n=120$ harmonic numbers in accordance with the energistate index range $(-120,120)$ in Fig. 2 (c) for correspondent number of atoms.   
As shown in Fig. 14, in contradistinction to the RGQD with an armchair edge, in the case of a zigzag edge on the elonged side of RGQD, if for the $y$--component 
we have maxima (red and black dots) in the high-frequency part of HHG corresponding to 
transitions from occupied states to unoccupied ones (the electron-hole 
creation and subsequent recombination), meanwhile for the $x $--component the preferable 
angles (in accordance with Fig. 5--8) appear in the low-frequency part of HHG spectra for the 
electron/hole transitions between nearest-neighbor eigenstates (within unoccupied/occupied states). 
Former transitions correspond to the interband current, which represents the recombination/creation of electron-hole pairs, 
while the latter corresponds to the intraband current. As shown in Fig. 14, that is defined by 
respective state energy difference ${n}\hbar\omega$. 
The interband current sufficiently enhances the HHG process rate in the elonged zigzag edge case.%
\begin{figure}[tbp]
\includegraphics[width=.76\textwidth]{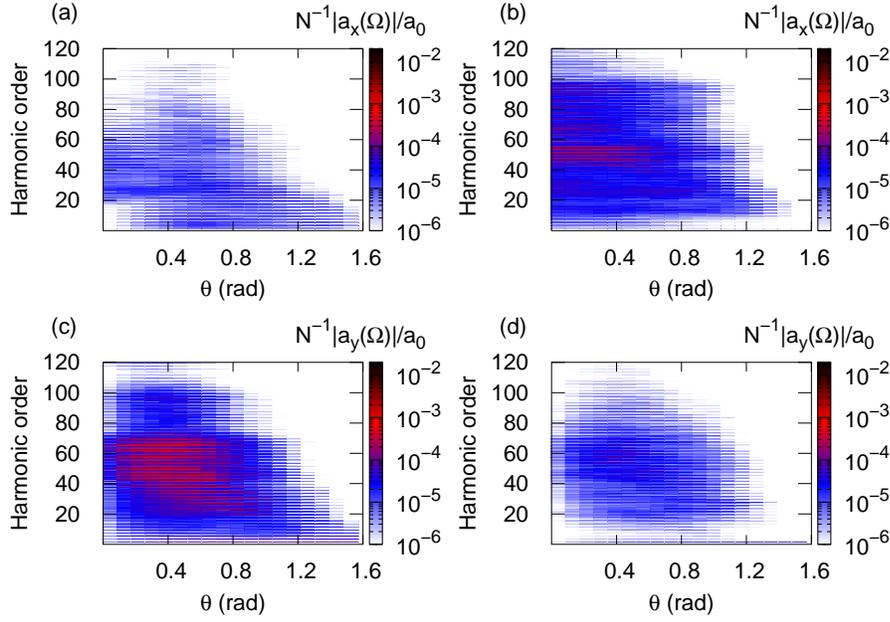}
\caption{(Color online) The color bars represent the HHG emission rate in
strong field regime in the logarithmic scale via dipole acceleration Fourier
transformation $N^{-1}|a_{i}\left( \Omega \right) |/a_{0}$ (in arbitrary
units, $ i=x,y$ ) versus $\theta$ and harmonic order for (a) $N=244$ at zigzag edge and (b) $N=252$ at armchair edge of RGQD. The wavefield
strength is $E_{0}=0.1\ \mathrm{V/\mathring{A}}$ with frequency $\protect\omega =0.1\ \mathrm{eV}/\hbar $, and the EEI energies
are $U=3\ \mathrm{eV}$ and $V\simeq 0.9\ \mathrm{eV}$. The relaxation rate
is taken $\hbar \protect\gamma =50\ \mathrm{meV}$. }
\end{figure}%

\section{Conclusion}

We have investigated the generation of infrared laser field harmonics in the 
extreme nonlinear strong field regime on 
the RGQD of different lateral sizes and edges. We treated the system evolution in the strong laser 
field on the basis of the Hartree-Fock approximation. The multiparticle Coulomb interaction 
we consider in the extended Hubbard approximation. The numerical results revealed 
the typical non-perturbative multiple-plateaus of the spectrum of HHG, the 
boundaries of which are determined by the internal excitation lines of the dot, 
and the differences related to the RGQD edge. Note, that the dominant plateau 
shifted towards higher frequencies with an increase in the number of atoms. 
Therefore, by changing the lateral size of the nanostructure, it is possible to increase 
the orders of harmonics within the main plateau. In addition, 
the cutoff photon energy also shifts toward blue with an increase in the transverse size 
of the nanostructure. We have shown that the HHG spectra have a strong 
anisotropy depending on the orientation of the electromagnetic wave field 
strength relative to the RGQD zigzag edge on the elonged side. Different polarization angles lead to different 
maxima in the harmonic spectra and cutoff energies. The isolated line corresponds to mainly
 an increase in the dipole momentum $y$--the component that appears in the HHG 
process on RGQD along the zigzag edge, which causes a sufficient increase in the high-order harmonics yield. The obtained 
results show that RGQD can serve as a medium for HHG at interacting with a strong laser 
field, for the appearance of new energy states, including that HHG probability increases 
with growth in the number of dot atoms. This can be a proper way to enhance the 
quantum yield of the HHG and the energy of photons in this process on the graphene-like quantum dots.%

\section{Appendixes: Interaction Hamiltonian}

Here we will present the total Hamiltonian by the empirical TB model \cite{Wal} in the form:%
\begin{equation}
\widehat{H}=\widehat{H}_{\mathrm{0}}+\widehat{H}_{\mathrm{int}},  \label{H1}
\end{equation}%
where  
\begin{equation}
\widehat{H}_{0}=\frac{1}{2}\sum_{\left\langle
i,j\right\rangle }V_{ij}n_{i}n_{j}+\frac{U}{2}\sum_{i\sigma }n_{i\sigma }n_{i%
\overline{\sigma }}-\sum_{\left\langle i,j\right\rangle \sigma
}t_{ij}c_{i\sigma }^{\dagger }c_{j\sigma }.  \label{Hfree}
\end{equation}%
is the free RGQD Hamiltonian.
Here $c_{i\sigma }^{\dagger }$ is the operator of creation of an electron with spin polarization $%
\sigma =({\uparrow ,\downarrow })$ at site $i$, and $\left\langle
i,j\right\rangle $ is the summation over nearest neighbor sites with
the transfer energy $t_{ij}$ ($\overline{\sigma }$ is the opposite to spin
polarization); and ${n}_{i\sigma }=c_{i\sigma }^{\dagger }c_{i\sigma }$ is the electron
density operator with total electron
density for the site $i$: ${n}_{i}={n}_{i\uparrow }+{n}_{i\downarrow }$. 
The first and the second terms in free Hamiltonian (\ref{Hfree}) correspond to the EEI
within the extended Hubbard approximation ($\widehat{H}_{ee}$) with
inter-site $\sim V_{ij}$ and on-site $\sim U$ Coulomb repulsion energies. The
inter-site Coulomb repulsion is described by the distance $d_{ij}$ between
the nearest-neighbour pairs varied over the system: $\sim
V_{ij}=Vd_{min}/d_{ij}$ ($d_{min}$ is the the minimal nearest-neighbor
distance). For the all calculations we have taken $V=0.3U$ \cite{GQD3}, \cite{GQD4}. 
The third term in (\ref{Hfree}) is the kinetic energy part of the TB Hamiltonian with tunneling matrix
element $t_{ij}$ neighboring sites. The hopping integral $t_{ij}$ between
the nearest-neighbor atoms of GQDs can be determined experimentally, and is
usually taken to be $t_{ij}=2.7\ \mathrm{eV}$ \cite{1}. 
Note, that we neglected the lattice vibrations in the Hamiltonian.

The laser-RGQD interaction is described in the length-gauge via the pure
scalar potential:%
\begin{equation*}
\widehat{H}_{\mathrm{int}}=e\sum_{i\sigma }\mathbf{r}_{i}\mathbf{E}\left(
t\right) c_{i\sigma }^{\dagger }c_{i\sigma },
\end{equation*}%
where $\mathbf{r}_{i}$ is the position vector, $e$ is the elementary charge. 
We will obtain evolutionary equations for
the single-particle density matrix $\rho _{ij}^{\left( \sigma \right)
}=\left\langle c_{j\sigma }^{\dagger }c_{i\sigma }\right\rangle $  from
the Heisenberg equation $i\hbar \partial \widehat{L}/\partial t=\left[ 
\widehat{L},\widehat{H}\right] $. Let us the system relaxes at a rate $\gamma $ to the equilibrium $%
\rho _{0ij}^{\left( \sigma \right) }$ distribution. The EEI will be considered under the Hartree-Fock
approximation, and to describe a closed set
of equations for the single-particle density matrix $\rho _{ij}^{\left(
\sigma \right) }$, we will assumed the Hamiltonian (\ref{Hfree}) in the form: 
\begin{equation*}
\widehat{H}_{0}^{HF}\simeq -\sum_{\left\langle i,j\right\rangle \sigma
}t_{ij}c_{i\sigma }^{\dagger }c_{j\sigma }+U\sum_{i}(\overline{n}_{i\uparrow
}-\overline{n}_{0i\uparrow })n_{i\downarrow }
\end{equation*}%
\begin{equation*}
+U\sum_{i\sigma }(\overline{n}_{i\downarrow }-\overline{n}_{0i\downarrow
})n_{i\uparrow }+\sum_{\left\langle i,j\right\rangle }V_{ij}(\overline{n}%
_{j}-\overline{n}_{0j})n_{i}
\end{equation*}%
\begin{equation}
-\sum_{\left\langle i,j\right\rangle \sigma }V_{ij}c_{i\sigma }^{\dagger
}c_{j\sigma }\left( {\left\langle {c_{i\sigma }^{\dagger }c_{j\sigma }}%
\right\rangle }-{\left\langle {c_{i\sigma }^{\dagger }c_{j\sigma }}%
\right\rangle }_{0}\right) ,  \label{HFU}
\end{equation}%
with $\overline{n}_{i\sigma }={\left\langle {c_{i\sigma }^{\dagger
}c_{i\sigma }}\right\rangle }=\rho _{ii}^{\left( \sigma \right) }$. Thus,
the following equation for the density matrix is obtained: 
\begin{equation*}
i\hbar \frac{\partial \rho _{ij}^{\left( \sigma \right) }}{\partial t}%
=\sum_{k}\left( \tau _{kj\sigma }\rho _{ik}^{\left( \sigma \right) }-\tau
_{ik\sigma }\rho _{kj}^{\left( \sigma \right) }\right) +\left( V_{i\sigma
}-V_{j\sigma }\right) \rho _{ij}^{\left( \sigma \right) }
\end{equation*}%
\begin{equation}
+e\mathbf{E}\left( t\right) \left( \mathbf{r}_{i}-\mathbf{r}_{j}\right) \rho
_{ij}^{\left( \sigma \right) }-i\hbar \gamma \left( \rho _{ij}^{\left(
\sigma \right) }-\rho _{0ij}^{\left( \sigma \right) }\right) ,  \label{evEqs}
\end{equation}%
and the matrixes $V_{i\sigma }$, $\tau _{ij\sigma }$ are approximated by the
density matrix $\partial \rho _{ij}^{\left( \sigma \right) }$: 
\begin{equation}
V_{i\sigma }=\sum_{j\alpha }V_{ij}\left( \rho _{jj}^{\left( \alpha \right)
}-\rho _{0jj}^{\left( \alpha \right) }\right) +U\left( \rho _{ii}^{\left( 
\overline{\sigma }\right) }-\rho _{0ii}^{\left( \overline{\sigma }\right)
}\right) ,
\end{equation}%
\begin{equation}
\tau _{ij\sigma }=t_{ij}+V_{ij}\left( \rho _{ji}^{\left( {\sigma }\right)
}-\rho _{0ji}^{\left( {\sigma }\right) }\right) .
\end{equation}%

In this representation, the initial value of density matrix ${\left\langle {%
c_{i\sigma }^{\dagger }c_{j\sigma }}\right\rangle }_{0}$ is defined via 
TB Hamiltonian ${\widehat{H}_{0}^{TB}}=-\sum_{\left\langle
i,j\right\rangle \sigma }t_{ij}c_{i\sigma }^{\dagger }c_{j\sigma }$. We will 
numerically diagonalize the TB Hamiltonian $\widehat{H}_{0}$. That is, in the
static limit the Hartree-Fock Hamiltonian vanishes $\widehat{H}%
_{ee}^{HF}\simeq 0$, and the EEI in Hartree-Fock limit is included in
empirical hopping integral between the nearest-neighbor atom $t_{ij}$ which
is chosen to be close to experimental data \cite{bb}. Thus, the EEI in the
Hartree-Fock approximation is correspond only for quantum dynamics initiated
by the pump laser field.

The authors are deeply grateful to prof. H. K. Avetissian and Dr. G. F. Mkrtchian for permanent discussions and valuable recommendations.
This work was supported by the Science Committee of RA in Frames of Project 21AG-1C014.

\end{document}